\documentclass[preprintnumbers,showpacs,amsmath,amssymb,floatfix,prd,onecolumn,superscriptaddress,nofootinbib]{revtex4}
\usepackage{graphicx}
\usepackage{epsfig}
\usepackage{bm}
\usepackage{amsfonts}
\usepackage{epstopdf}
\usepackage{bm}

\begin{document}

\title{Can be OPERA neutrino as tachyonic chameleon?}

\author{Chao-Jun Feng}
\affiliation{Shanghai United Center for Astrophysics (SUCA), \\ Shanghai Normal University,
    100 Guilin Road, Shanghai 200234, P.R.China}

\author{Xin-Zhou Li}
\email{kychz@shnu.edu.cn} \affiliation{Shanghai United Center for Astrophysics (SUCA),  \\ Shanghai Normal University,
    100 Guilin Road, Shanghai 200234, P.R.China}

\author{Dao-Jun Liu}
\affiliation{Shanghai United Center for Astrophysics (SUCA), \\ Shanghai Normal University,
    100 Guilin Road, Shanghai 200234, P.R.China}

\begin{abstract}
Whether there is a superluminal object existing in the real world is a mystery in modern physics. Recently, OPERA collaboration claims that they have observed superluminal neutrinos in their experiments. This leads to a rest mass paradox by combining the observational results from SN1987A, if one considers the neutrino as  a tachyon. In this paper, we identify the neutrino as a tachyonic chameleon, whose mass depends not only on coordinate frame but also on the environment, in which the neutrino propagates without the need of violation of the Lorentz-invariance at a fundamental level. We have compared our results with these observations and made some predictions that can be tested in the near future. Also, we have clarified the concept of rest mass for subluminal and superluminal object.

\end{abstract}


\maketitle


\section{Introduction}
There is an essential distinction between the concept of an invariant speed and that of a maximum speed, so  the principle of relativity does not repel the superluminal ($v^2 > c^2$) object. If suitable postulates are chosen, the theory of relativity only requires that there be an invariant speed, and then one could get a consistent kinematic theory. Our basic starting point is (e.g. see Ref~\cite{Recami:1984xh} and refs. therein) as the following: (i) The principle of relativity: The physical laws should be valid not only for a particular observer, but for all inertial observers; (ii) For each inertial observer, the space that he or she lives in appears as homogeneous and isotropic, and the time also appears as homogeneous. These postulates imply that there is one and only one  invariant speed $w$: if one assumes $w=\infty$, then one gets the theory of relativity with Galilean transformations; if one assumes $w$ was finite and real, and from the experiments that told us such a speed is the speed of light in vacuum, then one would get the theory of relativity with not only ordinary subluminal Lorentz transformations (LT), but also the  superluminal Lorentz transformations (SLT), which is an extended theory of the Einstein's special relativity. Superluminal and subluminal objects have been named tachyons and bradyons, which come from the greek word $\tau \alpha \chi {\acute{\upsilon}} \varsigma$ and $\beta \rho \alpha
\delta {\acute{\upsilon}} \varsigma$, see Ref.~\cite{Feinberg:1967zz}. At last, objects traveling exactly at the speed of light are called ``luxons".

Since the light speed is invariant, all inertial frames can be divided  into two complementary subset $\{s\}$ and $\{S\}$, frames of $\{s\}$ ($\{S\}$) having speeds  $|v| < c$ ($|v| > c$) relative to a fixed $s_0\in \{s\}$. The concepts of tachyon and bradyon, and $\{s\}$ and $\{S\}$ do not have an absolute, but only a relative meaning. Under LT from $s_1$ to $s_2$ ($s_i\in \{s\}$), or from $S_1$ to $S_2$ ($S_i\in\{S\}$), the $4$-vector type is preserved. However, SLT from $s_1$ to $S_2$, or from $S_1$ to $s_1$ must transform timelike vectors into spacelike vectors, and vice versa. The energy $E$ and momentum $\mathbf{p}$ forms a $4$-vector $p^\mu$. Set $p^\mu = (E, \mathbf{p})$ in $s_0\in\{s\} $ and $p^{'\mu} = (E', \mathbf{p}')$ in $S_0\in\{S\} $, then
\begin{equation}\label{massenergy}
  E^2 - p^2c^2 = p^{'2}c^2 - E^{'2} = - m_0^2 c^4 \,,
\end{equation}
where $m_0$ is the rest mass for the frames of $\{S\}$, therefore $m_0$ is real but not image. We shall consider ourselves as the observer $s_0$, so that we have
\begin{equation}\label{Li:mass-energy}
  \frac{v^2 - c^2}{c^2} = \frac{m_0^2c^4}{E^2} \,,
\end{equation}
where energy $E$ of a tachyon via its velocity $v>c$ in the frame $s_0$.

Van Dam et al. point out that the tachyons can not be realized in the framework of field theory \cite{vanDam:1985we}. However, their conclusion is based on the unitarity restriction: any particle in relativistic quantum theory must described by unitrary irreducible representations of Poincar\'e algebra or its supersymmetric generalization. If we abandon this restriction, then we can construct the supersymmetric invariant action for supersymmetric tachyons \cite{Li:1987sx}. In other words, we can construct action for tachyonic boson and fermions \cite{Li:1986xy}.

Recently, the OPERA collaboration \cite{OPERA:2011zb} has provided evidence of superluminal $\nu_\mu$ propagation between CERN and the LNGS with $(v^2-c^2)/c^2 \approx 5\times 10^{-5}$, conforming an earlier record from the MINOS detector \cite{Adamson:2007zzb}. Before drawing a final verdict, certainly, the most urgent thing is to further check systematic errors in the measurement. However, the old Fermilab and SN1987A results are consistent with luminal propagation \cite{Kalbfleisch:1979rm}. In the SN1987A case $|v-c|/c < 2\times 10^{-9}$. Applied data into Eq.~(\ref{Li:mass-energy}), it gives $m_0 \approx 120$MeV/$c^2$ for OPERA, and $m_0 \approx 0.32 $KeV/$c^2$ for SN1987A. The rest mass $m_0$ of OPERA neutrino conflicts obviously with the SN1987A bound. If we assume that both are tenable, then these observations lead to a rest mass paradox.

In fact, the environments surround the neutrinos  in the experiments or observations are different for e.g. OPERA and SN1987A, so it is very natural to take account of the affect of the environment on the mass of the neutrino, which is much like a tachyonic chameleon. From the observations, one can see that this affect is important in the experiment such as OPERA. In the next two sections, we will introduce the tachyonic neutrino and show how the chameleon mechanism works, after that, we will compare our model with the observation results and also give some predictions that could be tested by experiment in near future. In the last section, we give some discussions on our model and some insights on the essence of superluminal phenomenon.

\section{Tachyonic neutrino}

Undoubtedly, the reliability of OPERA measurement lays for the forthcoming experiments aimed at reproducing such results. However, one can develop some exotic models taken at face OPERA and SN1987A data. It is possible to solve the rest mass paradox using a mechanism of tachyonic chameleon. We shall consider the rest mass $m_\nu$ of neutrino is depending on the energy density of the environment $\rho$. In other words, $m_\nu$ should be a function of $\rho$, not a constant. Taken at face value, $m_\nu(\rho)$ is $120$MeV/$c^2$ for OPERA, and $ 0.32 $KeV/$c^2$ for SN1987A.

It is convenient to describe the tachyonic neutrino when we abandon the restriction of unitary representation
\begin{equation}\label{Li:lag}
  \mathcal{L}_\nu = \frac{i}{2} \, \bar\psi \gamma^{\mu} \overleftrightarrow{\partial_\mu}\psi - m_\nu(\rho) \bar\psi\gamma_5 \psi \,,
\end{equation}
in the natural unit. Here $\overleftrightarrow{\partial_\mu}$ is defined as $b \overleftrightarrow{\partial_\mu} a = b \partial_\mu a - (\partial_\mu b )a$, and
\begin{equation}
  \psi \equiv
        \left(
          \begin{array}{c}   
            \xi \\  
            \eta \\  
          \end{array}
        \right) \,,
\end{equation}
which is a spinor with four components.
The component $\xi\in F$, where $F$ is the two-dimensional complex representation space of SL(2,C), whose elements are so called left-handed Weyl spinors; $\eta\in \dot F^*$, where $\dot F^*$ is the two-dimensional complex representation space of SL(2,C), whose elements are so called right-handed Weyl spinors.

From the Lagrangian (\ref{Li:lag}), one can get the equation of motion straightforward as follows
\begin{eqnarray}
  i \frac{\partial\xi}{\partial t} &=& i \boldsymbol{\sigma}\cdot \nabla \xi + m_\nu(\rho) \, \eta \,, \\
  i \frac{\partial\eta}{\partial t} &=& -i \boldsymbol{\sigma}\cdot \nabla \eta - m_\nu(\rho) \, \xi \,.
\end{eqnarray}
For a free tachyonic neutrino, these equations have the  superluminal plane wave solution
\begin{equation}
  \xi\sim\eta\sim \exp\big[i(\mathbf{p}\cdot\mathbf{x} - Et)\big]\,.
\end{equation}
Also, from the Lagrangian (\ref{Li:lag}) and the following, one will see that in our model, the mass of neutrino is indirectly depending on the energy density of the environment $\rho = \rho_E$ through a scalar field coupled to the neutrino, whose value depends on $\rho_E$.

Recently, Dvali and Vikman~\cite{Dvali:2011mn} have suggested that OPERA results could be an environmental effect of the local neighborhood of our planet. However, their mechanism requires the existence of an exotic spin-2 field that is coupled to neutrinos and rest of the matter asymmetrically, both in the sign and in the magnitude. In this paper, $m_\nu(\rho)$ stems from the Yukawa interaction with the chameleon scalar~\cite{Khoury:2003rn,Khoury:2003aq}, in which we need not require the existence of new fields.

\section{Chameleon mechanism}

The chameleon mechanism of scalar field operates whenever a scalar field couples to matter in such a way that its effective mass depends upon the energy density $\rho_E$ of the environment~\cite{Khoury:2003rn, Khoury:2003aq,Gubser:2004uf, Upadhye:2006vi, Liu:2010qq}. This mechanism also has been developed in $f(R)$ theory~\cite{Carroll:2003wy}.

For simplicity, we consider only the situation that there is a single environmental matter component. In this case, the action takes the following form \cite{Khoury:2003rn,Khoury:2003aq}
\begin{eqnarray}\label{action:chameleon}
S&=&\int d^4x \sqrt{-g}\left[\frac{M_{pl}}{2}\mathcal{R}+\frac{1}{2}\partial_\mu\phi\partial^\mu\phi-V(\phi)\right]\nonumber\\
&-&\int d^4x\mathcal{L}_m\left[\Psi_m,\tilde{g}_{\mu\nu}\right],
\end{eqnarray}
where $M_{pl}\equiv (8\pi G)^{-1/2}$ is the reduced Planck mass and $\mathcal{R}$ is the Ricci scalar and $\Psi_m$ denotes matter field. The scalar field $\phi$ couples conformally to matter particles through a coupling of the form $e^{-\beta \phi/M_{pl}}$ where $\beta$ is a positive dimensionless constant. It is worth noting that  the sign of $\beta$ in our paper is different with that in Ref.~\cite{Dvali:2011mn, Khoury:2003rn, Khoury:2003aq,Gubser:2004uf, Upadhye:2006vi, Liu:2010qq},  Meanwhile, $\Psi_m$ couples to metric $\tilde{g}_{\mu\nu}$, which is obtained by a transformation from the Einstein-frame metric $g_{\mu\nu}$,
\begin{equation}
\tilde{g}_{\mu\nu}=e^{-2\beta \phi/M_{pl}}g_{\mu\nu}.
\end{equation}
For non-relativistic matter, the energy density is determined by $\tilde{\rho}=\tilde{g}^{\mu\nu}T_{\mu\nu}$, where $T_{\mu\nu}=(2/\sqrt{-\tilde{g}})\delta\mathcal{L}_m/\delta\tilde{g}^{\mu\nu}$ is the stress tensor.
From action (\ref{action:chameleon}), it is obtained that
$\nabla^2\phi={d V_{\mathrm{eff}}}/{d \phi}$, where the effective potential
\begin{equation}
V_{\mathrm{eff}}=V(\phi)+\rho e^{-\beta\phi/M_{pl}}.
\end{equation}
Here we have defined $\rho\equiv \tilde{\rho}e^{-3\beta\phi/M_{pl}}$, which is conserved in the Einstein frame and independent of scalar field $\phi$.
The key point for the dynamics of $\phi$ is that it depends not only on the self interaction $V(\phi)$, but also on the environment density $\rho_E$. Due to this property, $\phi$ is dubbed as chameleon field.

In order to let $V_{\mathrm{eff}}$ has a minimum, we assume $V(\phi)$ is monotonically increasing. Therefore, the value of $\phi$ at the minimum of $V_{\mathrm{eff}}$, $\phi_{\mathrm{min}}$, and the mass of the fluctuations around the minimum, $m_{\phi,\mathrm{eff}}$, are both dependent on $\rho$,
\begin{equation}
\frac{\beta \rho}{M_{pl}}=V'(\phi_{\mathrm{min}})e^{\beta\phi_{\mathrm{min}}/M_{pl}},
\end{equation}
\begin{equation}
m_{\phi}^2=V''(\phi_{\mathrm{min}})+\frac{\beta}{M_{pl}}V'(\phi_\mathrm{min}),
\end{equation}
where the primes denote derivatives with respect to $\phi$. It is easy to find that the larger the density of the environment, the larger the mass of the chameleon and the larger the value of $\phi_{\mathrm{min}}$.

As we have mentioned before, the mass of neutrinos is also dependent on the environment, that is, the mass term in spinor field can be written as
\begin{equation}
m_{\nu}(\rho)\bar{\psi}\gamma_5\psi.
\end{equation}

Recently, the dynamics of Brans-Dicke cosmology with varying mass fermions is investigated \cite{Liu:2010qq}, where a Yukawa-type coupling between spinor field and scalar field is considered. Here, for neutrinos, we may consider the Lagrangian
\begin{equation}
L_{\nu}=\frac{i}{2}\left[\bar{\psi}\Gamma^\mu D_{\mu}\psi-(D_\mu\bar{\psi})\Gamma^\mu\psi\right]-\lambda\phi\bar{\psi}\Gamma^5\psi,
\end{equation}
where $\Gamma^{\mu}=e^{\mu}_{a}\gamma^a$ is generalized Dirac-Pauli matrices, $D_{\mu}$ denotes covariant derivatives, which is defined by
\begin{equation}
D_{\mu}=\partial_{\mu}+\frac{g_{\mu\nu}}{4}\left[\Gamma^{\nu}_{\sigma\lambda}-e^{\nu}_{b}(\partial_{\sigma}e^{b}_{\lambda})\right]
 \gamma^{\sigma}\gamma^{\lambda},
\end{equation}
where $\Gamma^{\nu}_{\sigma\lambda}$ and $e^{\nu}_{b}$ denote the Christoffel symbol and the tetrad, respectively. The metric tensor $g_{\mu\nu}$ satisfies the relation
\begin{equation}
g_{\mu\nu}=e^a_{\mu}e^b_{\nu}\eta_{ab},
\end{equation}
where $\eta_{ab}$ is the Minkowski metric tensor. Then, it is naturally assumed that, when chameleon scalar is located at $\phi_\mathrm{min}$ which is determined by local environment,   $m_{\mu}(\rho)=\lambda \phi_{\mathrm{min}}$, where $\lambda >0$ is dimensionless constant.
\section{Observations and Predictions }\label{sec:op}
To compare with observations, we simply take the effective potential as
\begin{equation}\label{miniPot}
  V_{\mathrm{eff}}(\phi) = \frac{1}{2}u^2\phi^2 + \rho_E e^{-\beta \phi} \,,
\end{equation}
which has a minimal value at $\phi_{min}$ that satisfies
\begin{equation}\label{phimineq}
   u^2\phi_{min} = \beta \rho_E e^{-\beta \phi_{min}} \,,
\end{equation}
and the corresponding mass of the scalar field is given by
\begin{equation}\label{massofphi}
  m^2_{\phi} = V''_{\mathrm{eff}}|_{\phi_{min}} = u^2 + \rho_E \beta^2 e^{-\beta \phi_{min}} \,.
\end{equation}
By using the mass-energy relation, we get the velocity of the neutrino as
\begin{equation}\label{vofn}
  \frac{v-c}{c} = \sqrt{1+ \frac{\lambda^2 \phi_{min}^2 c^4}{E^2}} -1 \,,
\end{equation}
where we have used the $\lambda\phi_{min} = m_\nu$. Therefore, once the energy density of the environment $\rho_E$ and the energy of the neutrino are given, one can predict  the velocity of the neutrino by solving Eqs.~(\ref{phimineq}) and (\ref{vofn}), if one has had calibrated parameters $\beta$ and $\mu$ from the observations that already known. In the following, we will take the observational results from SN1987A and OPERA to determine parameters and make some predictions for future experiments.

Here we will assume that the Earth is modeled as a sphere of density $\rho^{(e)}_E = 10$g/cm$^3$, with an atmosphere $10$ km thick with $\rho^{(a)}_E=10^{-3}$g/cm$^3$. Also, we known that the baryonic gas and dark matter in our neighborhood of the Milky Way is $\rho^{(g)}_E \approx 10^{-24}$g/cm$^3$ and the critical energy density of the present universe $\rho_c \approx 10^{-29}$g/cm$^3$. Henceforth, $\phi^{(i)}_{min}$ and $ m^{(i)}_{\phi}, (i=e,a,g)$ denote the value of $\phi_{min}$  that minimizes $V_{eff}$ for $\rho_E = \rho^{(i)}_E$ and the corresponding masses of the scalar field, respectively.

By taking the  observation value of SN1987A ($\rho_E \approx \rho_E^{(g)}$) and OPERA ($\rho_E \approx \rho_E^{(a)}$), we can get
\begin{eqnarray}
  \beta &=& \frac{\lambda\ln \left(r_1\delta\right)}{m_\nu^{(a)}(1-\delta)} \approx 3.07\times 10^{19} \text{MeV}^{-1} \,, \\
  \mu^2   &=&\frac{\lambda r_2\rho_c}{m_\nu^{(a)}} \beta e^{-\beta m_\nu^{(a)}/\lambda}  \approx 2.56 \times 10^{15} \text{MeV}^2 \,.
\end{eqnarray}
where we have defined $r_1 = \rho_E^{(a)}/ \rho_E^{(g)} \approx 10^{21}$, $r_2 = \rho_E^{(a)}/ \rho_c \approx 10^{26}$ and $\delta = \phi_{min}^{(e)}/\phi_{min}^{(g)} \approx 10^{-5}$. Here we have used $m_\nu^{(a)} \approx 120$ MeV, $\rho_c \approx 10^{-32}$ MeV$^4$ and $\lambda\approx 10^{20}$ for precluding the existence of the fifth force and we will discuss it later.

\begin{figure}[h]
\begin{center}
\includegraphics[width=0.4\textwidth]{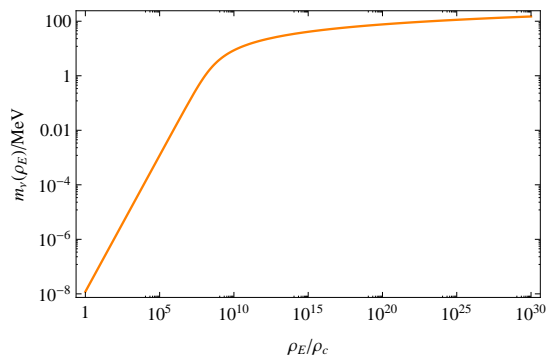}
\caption{\label{fig::mass}   The predicted neutrino mass with respect to the energy density of the environment.  }
\end{center}
\end{figure}

Since the neutrino mass $m_\nu$ is depending on the energy density of the environment $\rho_E$, it much looks like a chameleon. We have plotted the mass curve in Fig.~\ref{fig::mass}, which indicates that $m_\nu$ is increasing with $\rho_E$.

\begin{figure}[h]
\begin{center}
\includegraphics[width=0.4\textwidth]{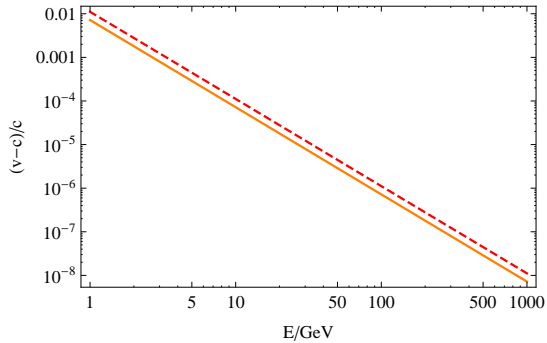}
\caption{\label{fig::vel}  The predicted velocity of the neutrino with respect to its energy in the environment of OPERA (Solid) and the core of the Earth (Dashed).  }
\end{center}
\end{figure}

In Fig.~\ref{fig::vel}, the  predicted velocity of neutrino is illustrated with respect of its energy $E$. Here, we would like to
propose a experiment that let the neutrino going through the Earth ($\rho_E \approx \rho_E^{(e)}$), and the corresponding theoretic prediction is given by the solid line in Fig.~\ref{fig::vel}. We also show the velocity curve of the neutrino in the OPERA environment.

The mass of the scalar field is given by Eq.~(\ref{massofphi}), one can easily get its value in the Earth as
\begin{equation}
  m_\phi^2 = \frac{r_2\lambda^2\rho_c}{m_\nu^2} \frac{\ln(r_1\delta)}{1-\delta} \left[1+ \frac{\ln(r_1\delta)}{1-\delta}\right] (r_1\delta)^{1/(\delta-1)} \,.
\end{equation}
By taking the observation value, one can get the mass corresponding to $\rho_E \approx \rho_E^{(a)}$ as $m_\phi\approx3.11\times10^{-12}\lambda$MeV. So, if one takes $\lambda \approx 10^{20}$ as we did before, $m_\phi \approx 10^5$TeV, then it avoids us to see the fifth force under present experiment conditions.

\section{Discussion}

To reconcile OPERA results with the spectacular explosion of SN1987A in the Large Magellanic Cloud, one possible approach would be to suggest violation of the Lorentz invariance at the fundamental level through some energy-dependent operators, see e.g. \cite{AmelinoCamelia:2011dx}. On the contrary, we suggested that the rest mass of neutrino depends on the environmental energy $rho_E$ without the need of violation of the Lorentz-invariance at a fundamental level.

It seems that once we identify the neutrino as a tachyonic chameleon, there would be no  rest mass paradox. In fact, the concept of rest mass is meaningful for subluminal objects, namely, these kind of objects in their rest frame have a well-defined rest mass. However, there is no well-defined 'rest' mass for superluminal objects, because the speed of a superluminal object can not cross the speed of light to become a subluminal object. Actually, if the speed of a superluminal object is much larger than that of light, its energy $E$ will be vanished because the contributions of the kinetic energy and the 'rest' mass are canceled each other. Thus, the 'rest' mass for a superluminal can be defined as the mass in  their infinite speed frame.

On the other side, from the LT and SLT transformation, one could get the velocity transformation for the frame $\{s\}$ and $\{S\}$ has a dual symmetry of $v \leftrightarrow c^2/v$. So, from this point of view, the rest frame for a subluminal object  and the infinite speed frame of a  superluminal object are dual to each other.

In our paper, the superluminal neutrino has a 'rest' mass defined as the mass in its infinite speed frame. Unfortunately,  we can not obtain its value from the mass-energy relation (\ref{massenergy}), but we can still get its effective mass from experiments or observations. We have compared our theoretic results with that from experiments, and made some predictions that can be tested in the near future. Since this is the first step to study  the neutrino as a tachyonic chameleon, there are still some unsolved problems, such as why the speed of light is the maximum (minimum) speed for a subluminal (superluminal) object. We will study these questions in further and it is hopeful that many experiments will be done to test this model.

\acknowledgments

This work is supported by National Science Foundation of China grant No.~11105091
and~11047138, National Education Foundation of China grant  No.~2009312711004, Shanghai Natural Science Foundation, China grant No.~10ZR1422000, Key Project of Chinese Ministry of Education grant, No.~211059,  and  Shanghai Special Education Foundation, No.~ssd10004.


\begin{thebibliography}{999}


\bibitem{Recami:1984xh}
  E.~Recami,
  Riv.\ Nuovo Cim.\  {\bf 9No 6}, 1 (1986).



\bibitem{Feinberg:1967zz}
  G.~Feinberg,
  Phys.\ Rev.\  {\bf 159}, 1089 (1967).



\bibitem{vanDam:1985we}
  H.~van Dam, Y.~J.~Ng and L.~C.~Biedenharn,
  Phys.\ Lett.\  B {\bf 158}, 227 (1985).


\bibitem{Li:1987sx}
  X.~Li and J.~Z.~Lu,
  J.\ Phys.\ A  {\bf 20}, 6113 (1987).


\bibitem{Li:1986xy}
  X.~Li and F.~Yu,
  Chin.\ Phys.\ Lett.\  {\bf 2}, 471 (1985).



\bibitem{OPERA:2011zb}
  T.~Adam {\it et al.}  [OPERA Collaboration],
  arXiv:1109.4897 [hep-ex].

\bibitem{Adamson:2007zzb}
  P.~Adamson {\it et al.}  [MINOS Collaboration],
  Phys.\ Rev.\  D {\bf 76} (2007) 072005
  [arXiv:0706.0437 [hep-ex]].


\bibitem{Kalbfleisch:1979rm}
  G.~R.~Kalbfleisch, N.~Baggett, E.~C.~Fowler and J.~Alspector,
  Phys.\ Rev.\ Lett.\  {\bf 43}, 1361 (1979).
  J.~Alspector {\it et al.},
  Phys.\ Rev.\ Lett.\  {\bf 36}, 837 (1976).
  K.~Hirata {\it et al.}  [KAMIOKANDE-II Collaboration],
  Phys.\ Rev.\ Lett.\  {\bf 58}, 1490 (1987).
  R.~M.~Bionta {\it et al.},
  Phys.\ Rev.\ Lett.\  {\bf 58}, 1494 (1987).


\bibitem{Dvali:2011mn}
  G.~Dvali and A.~Vikman,
  arXiv:1109.5685 [hep-ph].

\bibitem{Khoury:2003rn}
  J.~Khoury and A.~Weltman,
  Phys.\ Rev.\  D {\bf 69}, 044026 (2004)
  [arXiv:astro-ph/0309411].

\bibitem{Khoury:2003aq}
  J.~Khoury and A.~Weltman,
  Phys.\ Rev.\ Lett.\  {\bf 93}, 171104 (2004)
  [arXiv:astro-ph/0309300].

\bibitem{Gubser:2004uf}
  S.~S.~Gubser and J.~Khoury,
  Phys.\ Rev.\  D {\bf 70}, 104001 (2004)
  [arXiv:hep-ph/0405231].

\bibitem{Upadhye:2006vi}
  A.~Upadhye, S.~S.~Gubser and J.~Khoury,
  Phys.\ Rev.\  D {\bf 74}, 104024 (2006)
  [arXiv:hep-ph/0608186].



\bibitem{Liu:2010qq}
  D.~J.~Liu,
  Phys.\ Rev.\  D {\bf 82}, 063523 (2010)
  [arXiv:1005.5508 [astro-ph.CO]].



\bibitem{Carroll:2003wy}
  S.~M.~Carroll, V.~Duvvuri, M.~Trodden and M.~S.~Turner,
  Phys.\ Rev.\  D {\bf 70}, 043528 (2004)
  [arXiv:astro-ph/0306438].
  S.~Nojiri and S.~D.~Odintsov,
  Phys.\ Rev.\  D {\bf 68}, 123512 (2003)
  [arXiv:hep-th/0307288].

\bibitem{AmelinoCamelia:2011dx}
  G.~Amelino-Camelia, G.~Gubitosi, N.~Loret, F.~Mercati, G.~Rosati and P.~Lipari,
  arXiv:1109.5172 [hep-ph].






\end{thebibliography}
\end{document}